\documentclass[12pt]{article}
\newcommand{\phasecut}[1]{\int_{\vert{\mathbf #1}\vert<\Lambda} \frac{d^2#1}{(2\pi)^2}}
\newcommand{\phasecutabb}[1]{\int_{\vert{\mathbf #1}\vert<\Lambda}}
\newcommand{\phase}[1]{\int \frac{d^2#1}{(2\pi)^2}}

\newcommand{\create}[1]{a^\dagger_{#1}}
\newcommand{\destroy}[1]{a_{#1}}
\newcommand{\createvec}[1]{a^\dagger_{\mathbf #1}}
\newcommand{\destroyvec}[1]{a_{\mathbf #1}}
\newcommand{\angel}[1]{\chi^\dagger_#1}
\newcommand{\energy}[1]{\omega_{\mathbf #1}}
\newcommand{\energyman}[1]{\omega_{#1}}
\newcommand{\fourier}[2]{f_{#1}(#2)}
\newcommand{\fouriercon}[2]{f^*_{#1}(#2)}

\newcommand{\heatkernel}[2]{K_u(#1,#2)}
\newcommand{\Emin}{E_*}
\newcommand{\dij}{d_{ij}}
\newcommand{\dijmin}{d_{ij}^{min}}
\newcommand{\mumin}{\mu_i^{min}}
\newcommand{\coup}[1]{g_{#1}}
\usepackage{amssymb}

\begin{document}
\author{
\c{C}a$\bar{\rm g}$lar Do$\bar{\rm g}$an$^1$ , O. Teoman Turgut$^{1,2}$ 
\\ $^1$Feza G$\ddot{\rm u}$rsey Institute, \c{C}engelk\"{o}y, 34684 $\dot{\rm I}$stanbul, Turkey 
\\ $^2$Department of Physics, Bo$\bar{\rm g}$azi\c{c}i University, Bebek, 34342 $\dot{\rm I}$stanbul, Turkey}
\title{\bf Interaction of Relativistic Bosons with Localized Sources on Riemannian Surfaces}
\maketitle
\begin{abstract}
We study the interaction of mutually non-interacting Klein-Gordon particles with localized sources on stochastically complete Riemannian surfaces. This asymptotically free theory requires regularization and coupling constant renormalization. Renormalization is performed non-perturbatively using the orthofermion algebra technique and the principal operator $\Phi$ is found. The principal operator is then used to obtain the bound state spectrum, in terms of binding energies to single Dirac-delta function centers. The heat kernel method allows us to generalize this procedure to compact and Cartan-Hadamard type Riemannian manifolds. We make use of upper and lower bounds on the heat kernel to constrain the ground state energy from below thus confirming that our neglect of pair creation is justified for certain ranges of parameters in the problem.
\end{abstract}

\section{Introduction}
\label{introduction}
Quantum field theories only make sense after a regularization and renormalization procedure is applied to them. However, the complexity of quantum field theories, QCD in particular, combined with the lack of exact solutions in most cases obscures the concept of renormalization. It is extremely useful for understanding the effects of renormalization in QCD to investigate theories that are simple yet capture some of the important features of QCD such as asymptotic freedom and (classical) scale invariance.
\par From the above point of view, one would wish to understand renormalization in detail in the quantum mechanical context where the dynamics are much simpler. This has been done for the interaction of non-relativistic bosons with Dirac-delta function type potentials in dimensions 2 and 3 \cite{altunkaynak} where renormalization is required as well as for the relativistic Lee model \cite{kaynak}. These models all share the property of being asymtotically free. Furthermore, the interaction of non-relativistic bosons with Dirac-delta potentials in 2 dimensions, as the coupling constant is dimensionless, is a scale invariant theory classically.
\par Berezin and Fadeev \cite{berezin} were the first to investigate the interaction of non-relativistic bosons with Dirac-delta function potentials, this was then taken up in J. Hoppe's doctoral dissertation \cite{hoppe} in detail. Regularization and renormalization were performed by a scattering analysis \cite{jackiw, mitra, mead, perez, park, thorn} and by thinking of the Dirac-delta function as the limit of different potentials \cite{gosdzinsky, nyeo}, renormalization group properties of the problem were analyzed \cite{manuel, adhikari, adhikarighosh} and it was put in a broader mathematical framework in the theory of self-adjoint extensions of non-interacting hamiltonians \cite{albeverio}. The theory we investigate in this paper is a relativistic version of the interaction of non-relativistic bosons with Dirac-delta function potentials. We hope to get a bit closer to quantum field theory by allowing for a relativistic dispersion relation for the bosons even though we continue to neglect particle production due to the interaction with the source as the hamiltonian under scrutiny conserves the number of bosons. \par We use the principal operator technique introduced in an unpublished paper by Rajeev \cite{rajeev} to renormalize the theory nonperturbatively and to obtain its bound state spectrum. Then, along the lines of \cite{altunkaynak} we generalize the renormalization procedure to compact and Cartan-Hadamard type Riemannian manifolds in the hope of gaining some insight into how renormalization works and how it modifies the theory, if at all. We reach the conclusion that although the renormalization procedure that is crucial for obtaining physically meaningful results involves advanced mathematical techniques, it does not seem to contradict or even alter in any fundamental way our intuition from elementary quantum mechanics.
\par The paper is organized as follows: In section~\ref{generalformalism}, we write down the cut-off hamiltonian and explain how to obtain its spectrum through the eigenstates of the principal operator with vanishing eigenvalue. The principal operator is then renormalized in a way that is valid on Riemannian manifolds using the heat kernel. Subsection~\ref{sectioncouplingconstant} in section~\ref{renormalization} shows the equivalence between the renormalization done using the heat kernel and the usual method in field theory of placing a momentum space cut-off. The momentum space wavefunction of single particle fields and the corresponding configuration space ``wavefunctions'' are obtained in subsection~\ref{sectionwavefunction} of this section. The section concludes with an approximate calculation of bound state energies in the tunnelling regime in subsection~\ref{tunnellingregime}. Perhaps the most important results of the paper are the lower bounds we place on the ground state energy in section~\ref{groundbound} on flat, compact and Cartan-Hadamard type Riemannian manifolds.

\section{General Formalism}
\label{generalformalism}
We consider relativistic Klein-Gordon particles interacting with a finite number of localized sources on general two dimensional, stochastically complete Riemannian manifolds (${\mathcal M},g$). Our starting point is the following Hamiltonian
\begin{eqnarray}
\label{hamilton}
\nonumber &&H_\epsilon =H_0+H^{int}_\epsilon \\
\nonumber &&\mbox{where} \\
\nonumber &&H_0=\phase{p} \create{p} \destroy{p} \omega_p \\
&& H^{int}_\epsilon =-\sum_{i=1}^N \coup{i} \int d_gx K_{\epsilon/2}(a_i,x) \phi^{(-)}(x) \int d_gy K_{\epsilon/2}(a_i,y) \phi^{(+)}(y)
\end{eqnarray}
where $\coup{i}$ stands for the coupling constant. In the case of a non-relativistic particle interacting with Dirac-delta function type potentials in 1-dimension the coupling constant should be chosen positive (with the above convention) for bound states to exist. However, in our case, the sign of the coupling constant has little significance and can be of either sign depending on how it is renormalized. Here $N$ specifies the number of Dirac-delta function centers. The subscript $\epsilon$ indicates the regularization that is required to make the Hamiltonian above well-defined. In the $\epsilon \rightarrow 0^+$ limit that we are interested in, the regularized Hamiltonian has a Dirac-delta function type point interaction and the interaction above written in terms of the heat kernel $K_{\epsilon}(x,y)$ is just a smeared out version of it. A detailed discussion on this will be given later. The positive frequency part $\phi^{(-)}(x)$ of the bosonic field is given in terms of the creation operator indexed by $\sigma$ which has the interpretation of momentum vector on flat space as \\
\begin{eqnarray}
\label{phiminus}
\nonumber \phi^{(-)}(x)=\sum_{\sigma} \frac{\create{\sigma} \fouriercon{\sigma}{x}}{\sqrt{\energyman{\sigma}}} \\
\omega^2_{\sigma}=\lambda(\sigma)+m^2
\end{eqnarray}
\\ $\phi^{(+)}(x)$ is the Hermitian conjugate of this operator. The $f_{\sigma}(x) \in L^2({\mathcal M})$ are the orthonormal complete eigenfunctions of the Laplace-Beltrami operator which is defined on local coordinates $x \equiv (x^1,x^2)$ as
\begin{equation}
\label{laplacian}
\nabla^2_g \equiv \frac{1}{\sqrt{\mbox{det }g}} \sum_{\alpha,\beta=1}^2 \frac{\partial}{\partial x^{\alpha}}\left(g^{\alpha \beta} \sqrt{\mbox{det }g} \frac{\partial}{\partial x^{\beta}}\right)
\end{equation}
where $g_{\alpha \beta}$ are the elements of the metric tensor matrix $g$ and $g^{\alpha \beta}$ are those of its inverse. Eigenfunctions of the Laplace operator have the following properties
\begin{eqnarray}
\label{efunc}
\nonumber -\nabla_g^2 f_\sigma(x)=\lambda(\sigma) f_\sigma(x) \ , \\
\nonumber \int_{\mathcal M} d^{(n)}_g x f^*_\sigma(x) f_{\sigma^{\prime}}(x)=\delta_{\sigma \sigma^{\prime}} \ , \\
\sum_{\sigma} f^*_\sigma(x) f_\sigma(y)=\delta^{(n)}_g(x-y) \ .
\end{eqnarray}
The sums in the first line of Eq.~(\ref{phiminus}) and in the last relation of Eq.~(\ref{efunc}) should be replaced by integrals in the case of flat and Cartan-Hadamard type Riemannian manifolds since the spectrum of the Laplace operator on such manifolds is continuous. We should emphasize that whenever we talk about the eigenvalues of the Laplace operator we mean those of $-\nabla_g^2$ so that they are nonnegative.
\par This simple looking problem leads to divergences and some kind of regularization is mandatory to get physically sensible results. In order to renormalize the Hamiltonian in eq.~(\ref{hamilton}) we will use the orthofermion algebra technique developed by Rajeev \cite{rajeev}. In this method, we introduce fictitous operators $\chi$ and $\chi^\dagger$ named angels which are reminiscent of the hopping operators one encounters in solid state physics. These operators commute with the bosonic creation and annihilation operators and satisfy the relations given below
\begin{eqnarray}
\label{angels}
\nonumber \chi_i \angel{j}=\delta_{ij}\Pi_0 \ , \\
\nonumber \chi_i \chi_j=0 \ , \\
\sum_i \angel{i} \chi_i=\sum_i \Pi_i=\Pi_1 \ .
\end{eqnarray}
It is seen from the second relation and its Hermitian conjugate that these operators generate a two-state system when acting on the Fock space. $\Pi_0$ and $\Pi_i$ are projection operators onto the spaces with no angels and with one angel of type $i$ respectively. The physical space will be the direct product of the space of angels and the bosonic one. We define a new operator which in matrix form is as follows
\begin{eqnarray}
\label{newhamilton}
\nonumber G-E\Pi_0 &=& \left (
\begin{array}{cc}
(H_0-E) \Pi_0 & \sum_i \int d_gx K_{\epsilon/2}(a_i,x) \phi^{(-)}(x) \chi_i \\
\sum_j \int d_gy K_{\epsilon/2}(a_j,y) \phi^{(+)}(y) \angel{j} &  \sum_{k,l} \frac{\delta_{kl}}{\coup{k}} \angel{k} \chi_l \\
\end{array} \right ) \\
&\equiv &\left (
\begin{array}{cc}
a & b^\dagger \\
b & d \\
\end{array} \right )
\end{eqnarray}
with the resolvent or the Green's function defined as
\begin{equation}
 \label{resmatrix}
(G-E \Pi_0)^{-1} \equiv \left (
\begin{array}{cc}
\alpha & \beta^\dagger \\
\beta & \delta \\
\end{array} \right )
\end{equation}
The projection of this Green's function matrix on to the no angel subspace can be written in two alternative ways:
\begin{eqnarray}
\label{res}
\nonumber \alpha &=& (a-b^\dagger d^{-1} b)^{-1} = (H-E)^{-1} \\
&=&a^{-1}+a^{-1}b^\dagger \Phi^{-1}_{\epsilon} ba^{-1}
\end{eqnarray}
where the principal operator matrix $\Phi$ is given by
\begin{equation}
\label{character}
\Phi_{\epsilon} \equiv d-ba^{-1}b^\dagger
\end{equation}
The first relation in eq.~(\ref{res}) can easily be verified using the definition of $\Pi_0$ given in eq.~(\ref{angels}). This shows that the projection of the resolvent of the new operator onto the no angel subspace reproduces the Green's function of the original Hamiltonian. The reason we can project the Hamiltonian on to spaces with a fixed number of bosons is because it commutes with the number operator $N=\phase{p} \create{p} \destroy{p}$. The spectrum, i.e. the poles, of the original 
Hamiltonian corresponds to zeros of the principal operator matrix. The difficult task of finding the spectrum of the original Hamiltonian thus boils down to finding the eigenstates of this finite dimensional matrix $\Phi$ with eigenvalue zero.
\par When we use eq.~(\ref{character}) to find the explicit form of the principal operator matrix we get
\begin{eqnarray}
\label{charexp}
\nonumber \Phi_{\epsilon} &=& \sum_{i=1}^N \frac{1}{\coup{i}} \angel{i} \chi_i-\sum_{i=1}^N \int d_gx K_{\epsilon/2}(a_i,x) \sum_{\sigma} \frac{\destroy{\sigma}}{\sqrt{\energyman{\sigma}}} \fourier{\sigma}{x} \frac{\angel{i} \chi_j}{(H_0-E)} \\
&& \sum_{j=1}^N \int d_gy K_{\epsilon/2}(a_j,y) \sum_{\lambda} \frac{\create{\lambda}}{\sqrt{\energyman{\lambda}}} \fouriercon{\lambda}{y}
\end{eqnarray}
In the second term, angel operators can be moved around freely. However, the creation and annihilation operators do not commute with the free Hamiltonian and this expression is modified as we move the annihilation operator to the right of all the creation operators and vice versa. After normal-ordering we get
\begin{eqnarray}
\label{normalorder}
\nonumber \Phi_{\epsilon} &=& \sum_{i=1}^N \frac{1}{\coup{i}} \angel{i} \chi_i - \sum_{i,j} \sum_{\sigma} \int d_gx K_{\epsilon/2}(a_i,x) \fourier{\sigma}{x} \int d_gy K_{\epsilon/2}(a_j,y)\fouriercon{\sigma}{y} \\
\nonumber && \frac{1}{\energyman{\sigma}} \cdot \frac{\angel{i} \chi_j}{(H_0-E+\energyman{\sigma})} -\sum_{i,j} \int d_gx K_{\epsilon/2}(a_j,x) \sum_{\lambda}  \frac{\fouriercon{\lambda}{x} \create{\lambda}}{\sqrt{\energyman{\lambda}}} \\
&& \sum{\sigma} \frac{\angel{i}\chi_j}{(H_0-E+\energyman{\sigma}+\energyman{\lambda})} \int d_gy K_{\epsilon/2}(a_i,y) \frac{\fourier{\sigma}{y} \destroy{\sigma}}{\sqrt{\energyman{\sigma}}}
\end{eqnarray}
In the angel formulation, single boson states correspond to the ground state in the bosonic subspace. To illustrate this we feed in a 1-boson state to eq.~(\ref{res}).
\begin{equation}
\label{angelformulation}
\Phi_{\epsilon}^{-1}ba^{-1} \left(\vert 0> \bigotimes \vert p>_B\right) \propto \Phi_{\epsilon}^{-1}b \left(\vert 0> \bigotimes \vert p>_B\right)
\propto \Phi_{\epsilon}^{-1} \left(\sum_{i=1}^N c_i\vert \chi_i>\bigotimes \vert 0>_B\right)
\end{equation}
Besides, the application of the annihilation operator thus the free Hamiltonian on such states gives zero i.e.
\begin{equation}
\label{groundstate}
a_p \left (\sum_{i=1}^N c_i\vert \chi_i>\right )\bigotimes \vert 0>_B=\sum_{i=1}^N c_i\vert \chi_i> \bigotimes a_p\vert 0>_B=0
\end{equation}
Inspection of eq.~(\ref{normalorder}), in view of the above, shows that normal-ordering disentangles the effect of the bosonic operators and angels in the case of 1-particle states. In spite of the nonlinear realization of renormalization, it can be shown that the many particle states are given as direct products of the single particle ones just like in ordinary quantum mechanics. Therefore, knowledge of the spectrum of single particle states is equivalent to a knowledge of the full spectrum of the theory. This allows us to ignore the third term in eq.~(\ref{normalorder}) from now on.
\par We would like to write the principal operator in a form that is valid on Riemannian manifolds. We commence by separating the denominator using the identity below
\begin{equation}
\label{separatefrac}
\frac{1}{\energyman{p}(-E+\energyman{p})}=\frac{1}{(-E)} \left [ \frac{1}{\energyman{p}}-\frac{1}{(-E+\energyman{p})} \right ]
\end{equation}
followed by turning these denominators into exponentials with the help of the following relation:
\begin{eqnarray}
\label{trick}
\frac{1}{\energyman{p}}-\frac{1}{(-E+\energyman{p})} = \int_0^{\infty} e^{-s\energyman{p}} \left[1-e^{sE}\right] ds
\end{eqnarray}
This relation is valid provided the real parts of the terms in the exponentials satisfy $\Re \left(\energyman{p}-E \right)> 0$. One should think of the energy as a complex variable since the principal operator relevant for scattering can be obtained from this one by an appropriate analytical continuation in the complex energy plane.
\begin{eqnarray}
\label{normalordercont}
\nonumber \Phi_{\epsilon} &=& \sum_{i=1}^N \frac{1}{\coup{i}} \angel{i} \chi_i - \sum_{i,j} \sum_{\sigma} \int d_gx K_{\epsilon/2}(a_i,x) \fourier{\sigma}{x} \int d_gy K_{\epsilon/2}(a_j,y)\fouriercon{\sigma}{y} \\
&& \int_0^{\infty} ds \ e^{-s\energyman{\sigma}} \left[\frac{ 1-e^{sE}}{(-E)}\right] \angel{i} \chi_j
\end{eqnarray}
Another ingredient that we need is the subordination identity given below
\begin{equation}
\label{subordination}
e^{-sA}=\frac{s}{2\sqrt{\pi}} \int_0^{\infty} e^{-s^2/(4u)-uA^2} \frac{du}{u^{3/2}}
\end{equation}
Utilizing these relations in the expression for the principal operator we end up with the following expression
\begin{eqnarray}
\label{normalordersubordination}
\nonumber \Phi_{\epsilon} &=& \sum_{i=1}^N \frac{1}{\coup{i}} \angel{i} \chi_i - \frac{1}{2\sqrt{\pi}}\sum_{i,j} \int d_gx K_{\epsilon/2}(a_i,x) \int d_gy K_{\epsilon/2}(a_j,y) \int_0^{\infty} \frac{du}{u^{3/2}}\ e^{-um^2} \\
&&\sum_{\sigma} e^{-u\lambda(\sigma)}\fourier{\sigma}{x} \fouriercon{\sigma}{y} \int_0^{\infty} ds \ se^{-s^2/(4u)} \left[\frac{ 1-e^{sE}}{(-E)}\right] \angel{i} \chi_j \
\end{eqnarray}
We make a short digression to give information on the heat kernel $K_u(x,y)$, the minimal {\it positive} solution to the heat equation,
\begin{equation}
\label{heatequation}
\frac{\partial}{\partial u}\heatkernel{x}{y}=\nabla_g^2\heatkernel{x}{y}
\end{equation}
It has this eigenfunction expansion on compact Riemannian manifolds \cite{rosenberg, davies}
\begin{equation}
\label{efuncexpansion}
K_u(x,y)=\sum_{\sigma} e^{-u\lambda(\sigma)} \fourier{\sigma}{x} \fouriercon{\sigma}{y}
\end{equation}
and the following properties on general Riemannian manifolds
\begin{eqnarray}
\label{heatkernelproperties}
\nonumber \heatkernel{x}{y} =& \heatkernel{y}{x} & \mbox{symmetry property},\\
\nonumber \lim_{u \rightarrow 0^+} \heatkernel{x}{y} =& \delta_g(x,y) & \mbox{initial condition},\\
\int_{\mathcal M} d_gz \ K_{u_1}(x,z) \ K_{u_2}(z,y) =& K_{u_1+u_2}(x,y) & \mbox{reproducing property}.
\end{eqnarray}
We assume that the heat kernel satisfies the stochastic completeness relation given below
\begin{equation}
\label{stochasticcomp}
\int_{\mathcal M} K_u(x,y) d_gx = 1
\end{equation}
This is an immediate consequence of the eigenfunction expansion, however it will hold even when an eigenfunction expansion may be lacking. We identify the eigenfunction expansion of the heat kernel in eq.~(\ref{normalordersubordination}) and scale the $s$ variable to put it in the form below
\begin{eqnarray}
\label{normalorderheatkernel}
\nonumber \Phi_{\epsilon} &=& \sum_{i=1}^N \frac{1}{\coup{i}} \angel{i} \chi_i - \frac{1}{2\sqrt{\pi}}\sum_{i,j} \int d_gx K_{\epsilon/2}(a_i,x) \int d_gy K_{\epsilon/2}(a_j,y) \int_0^{\infty} \frac{du}{\sqrt{u}}\ e^{-um^2} \\
&& K_u(x,y) \int_0^{\infty} ds \ se^{-s^2/4} \left[\frac{ 1-e^{sE\sqrt{u}}}{(-E)}\right] \angel{i} \chi_j
\end{eqnarray}
We first use the symmetry property of the heat kernel,
\begin{eqnarray}
\label{normalorderreproducing}
\nonumber \Phi_{\epsilon} &=& \sum_{i=1}^N \frac{1}{\coup{i}} \angel{i} \chi_i - \frac{1}{2\sqrt{\pi}}\sum_{i,j}   \int_0^{\infty} \frac{du}{\sqrt{u}}\ e^{-um^2} \int_0^{\infty} ds \ se^{-s^2/4} \left[\frac{ 1-e^{sE\sqrt{u}}}{(-E)}\right] \\
&& \int d_gx \int d_gy K_{\epsilon/2}(a_i,x) K_u(x,y) K_{\epsilon/2}(y,a_j) \angel{i} \chi_j
\end{eqnarray}
then use the reproducing property and do an integration by parts to put the principal operator in the following form
\begin{eqnarray}
\label{normalorderintbyparts}
\Phi_{\epsilon} = \sum_{i=1}^N \frac{1}{\coup{i}} \angel{i} \chi_i - \frac{1}{\sqrt{\pi}}\sum_{i,j} \int_0^{\infty} \! ds \ e^{-s^2/4} \int_0^{\infty} \! du \ e^{sE\sqrt{u}} \ e^{-um^2} K_{u+\epsilon}(a_i,a_j) \angel{i} \chi_j
\end{eqnarray}
As we will discuss in more detail later, only the diagonal part of the principal operator leads to a divergence in the $\epsilon \rightarrow 0^+$ limit. We separate the divergent and finite parts of the principal operator and shift the $u$ integral keeping only the terms necessary to cut-off the divergence to write the principal operator as
\begin{eqnarray}
\label{normalordershift}
\nonumber \Phi_{\epsilon} &=& \sum_{i=1}^N \left[\frac{1}{\coup{i}} -\frac{1}{\sqrt{\pi}} \int_0^{\infty} ds \ e^{-s^2/4} \int_{\epsilon}^{\infty} du \ e^{sE\sqrt{u}} \ e^{-um^2} K_u(a_i,a_i)\right] \angel{i} \chi_i \\
&-& \frac{1}{\sqrt{\pi}}\sum_{i,j \atop (i\neq j)} \int_0^{\infty} ds \ e^{-s^2/4} \int_0^{\infty} du \ e^{sE\sqrt{u}} \ e^{-um^2} K_u(a_i,a_j) \angel{i} \chi_j
\end{eqnarray}
We choose the coupling constant in the following way so as to regularize the principal operator
\begin{eqnarray}
\label{couplingconstantmanifold}
\frac{1}{\coup{i}\left(\mu_i,\epsilon\right)} = \frac{1}{\sqrt{\pi}} \int_0^{\infty} ds \ e^{-s^2/4} \int_{\epsilon}^{\infty} du \ e^{s\mu_i\sqrt{u}} \ e^{-um^2} K_u(a_i,a_i)
\end{eqnarray}
This is a good place to make a digression and determine how the coupling constant changes with energy scale. In order to do this, we choose the renormalized coupling constant $g_{r,i}$ in terms of the bare coupling constant $g_{0,i}$ as
\begin{eqnarray}
\frac{1}{g_{r,i}(E,M_i)} = \frac{1}{g_{0,i}(\epsilon ,M_i)} -\frac{1}{\sqrt{\pi}} \int_0^{\infty} ds \ e^{-s^2/4} \int_{\epsilon}^{\infty} du \ e^{sE\sqrt{u}} \ e^{-um^2} K_u(a_i,a_i)
\end{eqnarray}
and impose the (unphysical) renormalization condition
\begin{eqnarray}
\frac{1}{g_{r,i}(E=M_i)}=0
\end{eqnarray}
The physically relevant case will be recovered if this renormalization scale $M_i$ is chosen to be the binding energy to the single $i^{\mbox{th}}$ Dirac-delta center. The condition above determines the bare coupling constant as a function of the renormalization scale and the renormalized coupling constant becomes the following:
\begin{eqnarray}
\frac{1}{g_{r,i}} = \frac{1}{\sqrt{\pi}} \int_0^{\infty} ds e^{-s^2/4} \int_0^{\infty} du \ e^{-um^2} K_u(a_i,a_i) \left[e^{M_i \sqrt{u} s}-e^{E\sqrt{u} s}\right]
\end{eqnarray}
The $i^{\mbox{th}}$ beta function $\beta_i(g_{r,i})$ is defined as follows:
\begin{eqnarray}
\beta_i(g_{r,i})=M_i \ \frac{\partial g_{r,i}}{\partial M_i}=-M_i g_{r,i}^2 \frac{\partial \left(1/g_{r,i}\right)}{\partial M_i}
\end{eqnarray}
Using this definition and imposing at the end the physical renormalization condition by setting $M_i=\mu_i$, we find the result given below for the beta function
\begin{eqnarray}
\beta_i(g_{r,i})=-\frac{\mu_i \ g_{r,i}^2(\mu_i)}{\sqrt{\pi}}\int_0^{\infty} s e^{-s^2/4} ds \int_0^{\infty} e^{-um^2} \sqrt{u} \ e^{\mu_i s\sqrt{u}} \ K_u(a_i,a_i) \ du
\end{eqnarray}
\par With the natural choice of the coupling constant as in eq.~(\ref{couplingconstantmanifold}), the experimentally measured binding energy $E$ coincides with the bound state energy $\mu_i$ in the single Dirac-delta center case. Inserting the expression for the coupling constant in eq.~(\ref{couplingconstantmanifold}) in the previous one yields the regularized form of the full principal operator:
\begin{eqnarray}
\label{charactermatrix}
\nonumber \Phi&=&\frac{1}{\sqrt{\pi}} \sum_{i=1}^N\int_0^{\infty} ds e^{-s^2/4} \int_0^{\infty} du e^{-um^2} K_u(a_i,a_i) \left[e^{\sqrt{u}\mu_i s}-e^{\sqrt{u}Es}\right]\angel{i} \chi_i \\
\nonumber &-& \frac{1}{\sqrt{\pi}} \sum_{i,j \atop (i \neq j)} \int_0^{\infty} ds e^{-s^2/4} \int_0^{\infty} du e^{-um^2} K_u(a_i,a_j) e^{\sqrt{u}Es}\angel{i} \chi_j \\
\nonumber &-& \frac{1}{4\pi} \sum_{i,j} \int_0^{\infty} ds s^2 \int_0^{\infty} \frac{du_1 du_2}{(u_1 u_2)^{3/2}} e^{-s^2\left(1/u_1+1/u_2\right)/4-m^2(u_1+u_2)} \\
&& \int d_gx \phi^{(-)}(x) K_{u_2}(x,a_j) e^{-s(H_0-E)} \int d_gy K_{u_1}(a_i,y) \phi^{(+)}(y) \angel{i} \chi_j
\end{eqnarray}
In the above equation, for completeness we included the term we had dropped earlier. As this expression is finite in the limit $\epsilon \rightarrow 0^+$, we dropped the subscript $\epsilon$ in $\Phi$. This is explained in more detail in eq.~(\ref{divergenceonman}) and the paragraph following it. 
\par What we have done so far applies to Riemannian manifolds in general. However, we will give up on generality for a while for the sake of exposing the results more clearly. As will be seen, the general features regarding the renormalization of the coupling constant, behavior of the off-diagonal elements at large separations of the Dirac-delta centers and how the elements of the principal operator matrix vary with bound state energy hold on Riemannian manifolds, too.

\section{Renormalization and Spectrum of the Theory}
\label{renormalization}
\subsection{Coupling Constant Renormalization on ${\mathbb R}^2$}
\label{sectioncouplingconstant}
In this section, we will illustrate how the regularization and renormalization procedure valid on Riemannian manifolds that we descibed in the previous section is in fact equivalent to the usual momentum cut-off prescription we encounter in quantum field theory. Identifying the $\fourier{p}{x}$'s with the usual Fourier exponents and replacing the heat kernels with the Dirac-delta functions, eq.~(\ref{normalorder}) becomes the following:
\begin{eqnarray}
\label{normalorderflat}
\nonumber \Phi_{\Lambda} &=& \sum_{i=1}^N \left [ \frac{1}{\coup{i}} - \phasecutabb{p} \frac{1}{\energy{p}} \cdot \frac{1}{(-E+\energy{p})} \right ] \angel{i} \chi_i - \sum_{i,j \atop (i \neq j)} \phasecutabb{p} e^{i{\mathbf p} \cdot ({\mathbf a}_i - {\mathbf a}_j)} \frac{1}{\energy{p}} \cdot \frac{1}{(-E+\energy{p})} \angel{i} \chi_j \\
&-& \sum_{i,j} \phasecutabb{p} \phasecutabb{k} \frac{e^{i{\mathbf p} \cdot {\mathbf a}_i - {\mathbf k} \cdot {\mathbf a}_j}}{\sqrt{\energy{p} \ \energy{k}}} \ \createvec{k} \frac{1}{(H_0-E+\energy{p}+\energy{k})} \destroyvec{p} \ \angel{i} \chi_j
\end{eqnarray}
It is obvious that the second term in the first line of the above expression diverges for $n \geq 2$. This divergence must be regularized by placing a cut-off $\Lambda$ in the upper limit of that integral. This removes the problem of infinities, however it raises a new difficulty that all physical observables depend on $\Lambda$. This is circumvented by letting the coupling constant depend on the cut-off momentum i.e. $g_i \equiv g_i(\Lambda)$ in such a way that all physical observables are independent of it in the limit $\Lambda \rightarrow \infty$. This is what in quantum field theory is commonly referred to as renormalization. We propose the simple form below for the coupling constant to ensure this independence.
\begin{equation}
\label{couplingconstant}
\frac{1}{\coup{i}(\mu_i,\Lambda)}=\phasecut{p} \ \frac{1}{\energy{p}} \cdot \frac{1}{(\energy{p}-\mu_i)}
\end{equation}
\par Let us now evaluate these integrals to get an explicit expression for the principal operator matrix on flat surfaces. First, we will evaluate the diagonal terms. Separating the denominators as in eq.~(\ref{separatefrac}) and using the fact that the integrands are rotationally invariant makes their calculation trivial. The result is (with no summation over repeated indices)
\begin{eqnarray}
\label{flatdiagint}
\nonumber \Phi_{ii} &=& \lim_{\Lambda \rightarrow \infty} \int_0^{2\pi} \frac{d\theta}{(2\pi)^2} \int_0^{\Lambda} pdp \left [ \frac{1}{\energy{p}(-\mu_i+\energy{p})} - \frac{1}{\energy{p}(-E+\energy{p})} \right ] \\
\nonumber &=& \lim_{\Lambda \rightarrow \infty} \frac{1}{2\pi}\left [ \ln \left( \frac{\sqrt{\Lambda^2+m^2}-\mu_i}{m-\mu_i} \right )- \ln \left ( \frac{\sqrt{\Lambda^2+m^2}-E}{m-E} \right ) \right ] \\
&=& \frac{1}{2\pi}\ln \left ( \frac{m-E}{m-\mu_i} \right ) + {\mathcal O}(1/\Lambda)
\end{eqnarray}
The choice eq.~(\ref{couplingconstant}) for the coupling constant in some sense absorbs the cut-off momentum dependence and the final result given above for the diagonal elements are, as promised, independent of this cut-off momentum in the limit $\Lambda \rightarrow \infty$. As will be shown momentarily, off-diagonal matrix elements are already finite so the cut-off momentum dependence has successfully been removed.
\par Interestingly enough, the elements of the principal operator matrix that are now finite do not depend on the coupling constant! The dimensionless coupling constant $\coup{i}$ has been traded for a dimensionful parameter $\mu_i$ which is the experimentally measured binding energy to the $i^{\mbox{th}}$ Dirac-delta function center. This is the analog of dimensional transmutation in the field theory context.
\par Let us now calculate the off-diagonal elements. We use the trick of separating the denominator of two terms into two denominators, followed by turning these fractions into integrals over some variable as in eq.~(\ref{trick})
\begin{eqnarray}
\label{generalizeoffdiag}
\Phi_{ij}=\frac{-1}{2\sqrt{\pi}(-E)} \int_0^{\infty} s\left [1-e^{sE}\right ] ds \int_0^{\infty} \frac{du}{u^{3/2}} e^{-s^2/(4u)-um^2} \int_{\vec p} e^{-up^2+i{\mathbf p}\cdot {\mathbf d}_{ij}}
\end{eqnarray}
The separation vector is defined as ${\mathbf d}_{ij}\equiv {\mathbf a}_i-{\mathbf a}_j$. After grinding through the algebra we arrive at the simple result given below:
\begin{equation}
\label{flatoffdiagint}
\Phi_{ij}=\frac {-1}{2\pi}\int_0^\infty e^{-d_{ij}\left [m(s^2+1)^{1/2}-Es\right ]} \frac{ds}{(s^2+1)^{1/2}}
\end{equation}
where $d_{ij}$ is shorthand for the length of the separation vector. With a little effort, the diagonal elements of the principal operator can alternatively be obtained starting from the equation above. For this purpose, one may find useful the following expansion of the gamma function near its pole at zero.
\begin{eqnarray}
\label{gamma}
\lim_{n \rightarrow 0} \int_0^{\infty} x^{(n-1)} e^{-ax} dx = \lim_{n \rightarrow 0} a^{-n} \Gamma(n) \simeq \frac{1}{n} - \ln{a} -\gamma + {\mathcal O}(n)
\end{eqnarray}
\par Concentrating on states that contain only a single boson let us make some observations based on the results for the elements of the principal operator matrix. First of all, the divergence in the coupling constant is a mild, logarithmic one as one would expect from naive dimensional analysis since the coupling constant is dimensionless. Moreover, in the limit the minimum separation between the different Dirac-delta function centers $d_{ij}^{min}$ is much greater than the Compton wavelength of the particle that is $md_{ij}^{min}\gg 1$, off-diagonal elements of the principal operator matrix are much smaller than the diagonal elements and can be considered as a perturbation. In order to make this explicit, we separate the principal operator matrix into a diagonal $\Phi_D(E)$ and an off-diagonal $\delta \Phi(E)$ part as follows:
\begin{equation}
\Phi (E)=\Phi_D (E)+\delta\Phi (E)
\end{equation}
Then we look for solutions of the form
\begin{equation}
\vert v_i>=\vert v_i^0>+\vert\delta v_i>
\end{equation}
with $\vert v_i^0>$ satisfying
\begin{equation}
\Phi_D(E=\mu_i)\vert v_i^0>=0
\end{equation}
When we do this we find that the variation in the binding energies due to the presence of other Dirac-delta function centers is given by the equation below
\begin{equation}
\label{perturbative}
\delta E_i=E_i-\mu_i\simeq \left (\frac{\partial\Phi_{ii}}{\partial E}\vert_{E=\mu_i}\right )^{-1}\sum_{j\neq i}\frac{\vert \Phi_{ij}(\mu_i)\vert^2}{\Phi_{jj}(\mu_i)}
\end{equation}
This equation is reminiscent of the 2$^{nd}$ order non-degenerate perturbation theory result, however unlike that equation this one is correct nonperturbatively, too. Although the change in the binding energy to a single Dirac-delta function center is assumed to be a perturbation i.e. $\left \vert \delta E_i \right \vert \ll \left \vert \mu_i \right \vert$, nothing prohibits the initial binding energy itself from being nonperturbatively strong that is $m-\mu_i \sim m$. The above equation is valid only if there is no degeneracy in the initial binding energies. A similar equation can be derived for the degenerate case.
\par It is possible, in this regime, to evaluate the off-diagonal elements approximately by hand which becomes exact in the limit. One can then plug this result in eq.~(\ref{perturbative}) to find the change in the binding energy. For this purpose, we may start with eq.~(\ref{flatoffdiagint}), however this way of finding the approximate form of the off-diagonal elements does not lend itself to generalization to Riemannian manifolds. Instead, we will use the form of the principal operator which we derived in eq.~(\ref{charactermatrix}) that is valid on Riemannian manifolds. Here we just state the results that will be derived in subsection~\ref{tunnellingregime}.
\begin{eqnarray}
\label{tunnelflat}
\Phi_{ij}(E=\mu_i) = -\frac{1}{\sqrt{2\pi}} \cdot \frac{e^{-m\dij \sqrt{1-(\mu_i/m)^2}}}{\left(m\dij \sqrt{1-(\mu_i/m)^2}\right)^{1/2}} \times \left \{ \begin{array}{lr} \frac{1}{2} & \left\vert \mu_i/m \right\vert \ll (m\dij)^{-1/2} \\ 1 & (m\dij)^{-1/2} \ll \mu_i/m \ll 1-(m\dij)^{-1} \end{array} \right.
\end{eqnarray}
and
\begin{eqnarray}
\label{negenergytunnelflat}
\Phi_{ij}(E=\mu_i) = -\frac{1}{2\pi} \cdot \frac{e^{-m\dij}}{m\dij \vert \mu_i/m \vert} & & (m\dij)^{-1/2} \ll -\mu_i/m < 1
\end{eqnarray}
For example, in the double Dirac-delta center case the binding energies $\mu_1 > \mu_2 > 0$ are modified as follows:
\begin{eqnarray}
\label{tunnelexample}
\nonumber \frac{E_1}{m} \simeq \frac{\mu_1}{m} + \frac{1}{2\pi} \cdot \sqrt{\frac{m-\mu_1}{m+\mu_1}} \cdot \frac{\exp{\left(-2md_{12}\sqrt{1-(\mu_1/m)^2}\right)}}{m d_{12} \ln \left(\frac{m-\mu_2}{m-\mu_1}\right)} \\
\frac{E_2}{m} \simeq \frac{\mu_2}{m} - \frac{1}{2\pi} \cdot \sqrt{\frac{m-\mu_2}{m+\mu_2}} \cdot \frac{\exp{\left(-2md_{12}\sqrt{1-(\mu_2/m)^2}\right)}}{m d_{12} \ln \left(\frac{m-\mu_2}{m-\mu_1}\right)}
\end{eqnarray}
\par It should also be noted that whereas the absolute value of the off-diagonal elements decrease with decreasing energy, the on-diagonal elements increase with decreasing energy. This observation will play an important role when we look for a lower bound on the ground state energy. It turns out that all of these properties of the principal operator matrix hold on Riemannian manifolds as well.

\subsection{Renormalization on Riemannian Manifolds}
\label{riemannianmanifolds}
The renormalized principal operator appropriate for 1-particle states written in terms of the heat kernel after normal-ordering is given below
\begin{eqnarray}
\label{regularphi}
\nonumber \Phi &=& \frac{1}{\sqrt{\pi}} \sum_{i=1}^N \int_0^{\infty} ds \ e^{-s^2/4} \int_0^{\infty} du \ \left[e^{s\mu_i \sqrt{u}}-e^{sE\sqrt{u}}\right] \ e^{-um^2} K_u(a_i,a_i) \angel{i} \chi_i \\
&-& \frac{1}{\sqrt{\pi}} \sum_{i,j \atop (i\neq j)} \int_0^{\infty} ds \ e^{-s^2/4} \int_0^{\infty} du \ e^{sE\sqrt{u}} \ e^{-um^2} K_u(a_i,a_j) \angel{i} \chi_j
\end{eqnarray}
\par Superficially, the $s$-integral in the first two terms of the above equation looks like it might be divergent at large values of $u$, however, this suspicion is groundless. One can convince oneself that this is the case by completing the square and shifting the $s$-integral. Eq.~(\ref{efuncexpansion}) implies that the asymptotic form of the heat kernel at late times\footnote{We call $u$ in $K_u(x,y)$ the time variable, however its inverse has mass dimension 2.} is given by
\begin{equation}
\label{heatkernellatetimes}
\lim_{u\rightarrow \infty} \heatkernel{x}{y} \propto e^{-u\lambda_{min}}
\end{equation}
where $\lambda_{min}$ is the minimum eigenvalue of the Laplacian on the Riemannian manifold of interest. This can be seen from the eigenfunction expansion, however it may hold even when an eigenfunction expansion does not exist. We assume this to be the case in the following discussion. In general, when we talk about the mass of the particle what we mean is $\sqrt{m^2+\lambda_{min}}$. Since the Laplacian is a positive non-definite operator $\lambda_{min} \geq 0$ on $L^2({\mathcal M})$ integrable functions, in order for the $u$-integral thus the matrix elements to be finite, energy eigenvalues are restricted to the following range
\begin{equation}
\label{energyrestriction}
-m < E < +m
\end{equation}
The upper bound is trivially satisfied for an attractive interaction as is the case here. The lower bound, though, is nontrivial and indicates when our approximation of neglecting pair creation can no longer be trusted. If the interaction is strong enough to lower the energy by 2$m$ from its non-interacting value of $H_0+m$ then pairs will inevitably form. This is because the minimum energy needed to create a pair of bosons out of the vacuum is 2$m$. One would expect this approximation to break down earlier since the bosons, due to the absence of conserved quantum numbers, need not pop out of the vacuum in pairs.
\par The lower limit of the $u$-integral is also finite for distinct points since the heat kernel approaches a delta function at early times. This is precisely the reason why the on-diagonal elements of the principal operator matrix diverge. The well-known short time expansion of the heat kernel \cite{rosenberg} is given below
\begin{equation}
\label{heatkernelshorttime}
\lim_{u\rightarrow 0^+}\heatkernel{x}{x}\simeq \frac{1}{(4\pi u)^{n/2}}\sum_{k=0}^\infty c_k(x) u^k
\end{equation}
where the coefficient functions $c_k(x)$ are given in terms of the Riemann curvature tensor and the covariant derivatives evaluated at the point $x$, with $c_0(x)=1$. In our case of $n=2$, it is clear that only the first term will lead to a divergence and the divergence will be a logarithmic one.
\begin{eqnarray}
\label{divergenceonman}
\nonumber \Phi_{ii} &=& \lim_{\epsilon \rightarrow 0^+} \frac{-1}{\sqrt{\pi}} \int_0^\infty e^{-s^2/4} ds \int_{\epsilon^2}^{t^2_*} e^{-um^2+Es\sqrt{u}} K_u(a_i,a_i) du \\ 
\nonumber &\simeq& \lim_{\epsilon \rightarrow 0^+} \frac{-1}{\sqrt{\pi}} \int_0^\infty e^{-s^2/4} ds \int_{\epsilon^2}^{t_*^2} \frac{du}{4\pi u} \\
&=& \lim_{\epsilon \rightarrow 0^+} \frac{-1}{2\pi} \ln\left(\frac{1/\epsilon}{1/t_*}\right)
\end{eqnarray}
We see that the divergence structure is the same as what we found on $R^2$ in eq.~(\ref{flatdiagint}) if we identify the inverse of the time cut-off with the momentum cut-off $1/\epsilon \equiv \Lambda$ and take the inverse of the upper limit $t_*$ at which the small time approximation breaks down to be roughly $1/t_* \sim m-E$. The correspondence between the divergence structures is to be expected as renormalization modifies the small distance behavior of the theory and that is exactly when every Riemannian manifold looks like flat space. Addition of the first term in the brackets in the first line of eq.~(\ref{regularphi}) removes this divergence.
\par We utilized the orthofermion algebra technique to transform the original problem into one where we look for eigenstates of the principal operator matrix with vanishing eigenvalue. It was by no means necessary to use this technique, in fact we will now find the explicit form of the single particle state wavefunctions using a direct approach.

\subsection{Single Particle State Wavefunctions}
\label{sectionwavefunction}
 The 1-particle field satisfies the following operator equation
\begin{equation}
\label{schrodinger}
H \vert \Psi>=E\vert \Psi>
\end{equation}
with $H$ given by the $\epsilon \rightarrow 0^+$ limit of eq.~(\ref{hamilton}). The free part of the Hamiltonian suggests that we look for a solution of the form
\begin{equation}
\label{ansatz}
\vert \Psi>=\int_q \create{q} \tilde{\psi}(q)\vert 0>
\end{equation}
Plugging the 1-particle field ansatz above into the Hamiltonian of eq.~(\ref{hamilton}) and grouping terms we find that the momentum space wavefunction must satisfy the following equation
\begin{eqnarray}
\label{wavefunceq}
&& \nonumber \tilde{\psi}(p)=\frac{C^{-1/2}}{\sqrt \energyman{p} (\energyman{p}-E)} \sum_{j=1}^N \coup{j} \fourier{p}{a_j} \psi(a_j) \\
&& \nonumber \mbox{where} \\
&& \psi(a_j)=\int_q \frac{\fouriercon{q}{a_j}}{\sqrt{\energyman{q}}} \tilde{\psi}(q)
\end{eqnarray}
Here, $C$ is a normalization constant which will be determined later. Expressing $\psi(a_i)$ in terms of this momentum space wavefunction leads to the following consistency relation among the values of the wavefunction\footnote{Even though we loosely call this the wavefunction, it should be kept in mind that it is not the generalized Fourier conjugate of the momentum space wavefunction because of the factor $\sqrt{\energyman{q}}$ in the denominator in the second line of eq.~(\ref{wavefunceq}).} at the positions of the Dirac-deltas
\begin{equation}
\label{consistencyeq}
\left [\frac{1}{\coup{i}}-\int_p \frac{\left \vert \fourier{p}{a_i} \right \vert^2}{\energyman{p} (\energyman{p}-E)}\right ]\psi (a_i)-\left [\sum_{j=1 \atop (j \neq i)}^N \frac{\coup{j}}{\coup{i}}\int_p \frac{\fourier{p}{a_j} \fouriercon{p}{a_i}}{\energyman{p} (\energyman{p}-E)}\right ]\psi(a_j)=0
\end{equation}
This set of equations relate the values of the wavefunction at the position of one Dirac-delta to the others. This equation can be written in matrix form as follows:
\begin{equation}
\label{characteristiceq}
\Phi_{ij} \psi(a_j)=0
\end{equation}
This is precisely the matrix form of the principal operator appropriate for a single boson that we encountered in eq.~(\ref{charactermatrix}). This equation does not fix the values of the wavefunction at the position of all Dirac-delta centers, rather it determines all the ratios. Therefore, we prefer instead to use the following ratios when writing down the expression for the wavefunction
\begin{eqnarray}
\label{characteristicratio}
R(a_j) \equiv \frac{\psi(a_j)}{\left[\sum_k \left \vert \psi(a_k) \right \vert^2\right]^{1/2}}
\end{eqnarray}
As we will see, the wavefunction diverges at the positions of the Dirac-delta centers, however the above ratios are finite. In order for this set of equations to have solutions, the determinant of the principal operator matrix must be zero. This will be the case only for special values of the binding energy $E$. There will generically be $N$ different values of the binding energy for which this is the case.
\par Normalization of the single particle state wavefunctions determines the constant $C$ in the momentum space wavefunction.
\begin{eqnarray}
\label{wavefuncnormalize}
\nonumber <\Psi\vert \Psi> &=& <0\left \vert \int_p a_p \tilde{\psi}^*(p) \int_q \create{q} \tilde{\psi}(q) \right \vert 0> \\
\nonumber &=& \int_p \left \vert \tilde{\psi}(p) \right \vert ^2 \\
&=& C^{-1} \sum_{i,j} \coup{i} \coup{j} \psi^*(a_i) \psi(a_j) \int_p \frac{\fourier{p}{a_j} \fouriercon{p}{a_i}}{\energyman{p} \left( \energyman{p}-E\right)^2} =1
\end{eqnarray}
When written in terms of the heat kernel the above expression yields
\begin{equation}
\label{normconstant}
C=\frac{1}{\sqrt{\pi}} \sum_{i,j} \coup{i} \coup{j} \psi^*(a_i) \psi(a_j) \int_0^{\infty} \! \! ds \ e^{-s^2/4} \int_0^{\infty} \! \! du \sqrt{u} \ e^{-um^2} e^{sE\sqrt{u}} K_u(a_j,a_i)
\end{equation}
Incidentally, this equation shows that if the same problem were solved in 3 or more dimensions, a wavefunction renormalization would also be necessary. The ultraviolet behavior is worse than what Altunkaynak, Erman and one of us \cite{altunkaynak} found for the non-relativistic case where the wavefunction normalization is necessary only in dimensions 4 and higher. When we write the wavefunction in configuration space we find
\begin{equation}
\label{wavefunc}
\psi(x)=C^{-1/2} \sum_{i=1}^N \coup{i} \psi(a_i) \int_p \frac{\fourier{p}{a_i} \fouriercon{p}{x}}{\energyman{p} (\energyman{p}-E)}
\end{equation}
Using the fact that
\begin{eqnarray}
\label{couplingratio}
\lim_{\epsilon \rightarrow 0^+} \frac{\coup{i}(\epsilon)}{\coup{j}(\epsilon)}=1
\end{eqnarray}
the wavefunction can be written in terms of the heat kernel as follows: 
\begin{eqnarray}
\label{wavefuncheatkernel}
\nonumber \psi(x)&=&\pi^{-1/4}\sum_{i=1}^N R(a_i) \int_0^{\infty} ds \ e^{-s^2/4} \int_0^{\infty} du \ e^{-um^2} e^{sE\sqrt{u}} K_u(a_i,x) \\
&& \times \left[\sum_{j,k} R^*(a_j) R(a_k) \int_0^{\infty} ds \ e^{-s^2/4} \int_0^{\infty} du \sqrt{u} \ e^{-um^2} e^{sE\sqrt{u}} K_u(a_k,a_j) \right]^{-1/2}
\end{eqnarray}
\par It is worthwhile to see what the expectation values of the relativistic and interaction energies are for these 1-particle states. For simplicity, let's calculate the relativistic energy carried by these 1-particle fields for the single Dirac-delta function center case.
\begin{eqnarray}
\label{kineticenergy}
\nonumber <\Psi \vert H_0 \Psi> &=& <0\vert \left (\int_p \destroy{p} \tilde{\psi}^\dagger(p) \right ) \left( \int_q \omega_q \tilde{\psi}(q) \create{q} \right )\vert 0> \\
\nonumber &=& C^{-1} g^4(\mu_i) \left \vert \psi(a_i) \right \vert^2 \int_p \frac{\left \vert \fourier{p}{a_i} \right \vert^2}{\left(\energyman{p}-E \right)^2} \\
&=& \int_p \frac{\left \vert \fourier{p}{a_i} \right \vert^2}{\left(\energyman{p}-E \right)^2} \left(\int_q \frac{\left \vert \fourier{q}{a_i} \right \vert^2}{\energyman{q} \left(\energyman{q}-E \right)^2}\right)^{-1}
\end{eqnarray}
With the use of the heat kernel one can convert the expression for the relativistic energy into the form below:
\begin{eqnarray}
\label{kinenergyheatkernel}
\nonumber <\Psi \vert H_0 \Psi>&=&\int_0^{\infty} ds \ s^2 e^{-s^2/4} \int_0^{\infty} du \ e^{-um^2} e^{sE\sqrt{u}} K_u(a_i,a_i) \\ 
&& \times \left[\int_0^{\infty} ds \ e^{-s^2/4} \int_0^{\infty} du \sqrt{u} \ e^{-um^2} e^{sE\sqrt{u}} K_u(a_i,a_i)\right]^{-1}
\end{eqnarray}
The relativistic energy carried by the field is infinite! Let's also calculate the energy the field has due to the interaction.
\begin{eqnarray}
\label{potentialenergy}
\nonumber <\Psi \vert H^{int} \Psi> &=& -g^2(\mu_i) \psi^*(a_i) \psi(a_i) \\
\nonumber &=& - C^{-1} g^6(\mu_i) \left \vert \psi(a_i) \right \vert^2 \int_p \frac{\left \vert \fourier{p}{a_i} \right \vert^2}{\energyman{p} \left(\energyman{p}-E\right)} \int_q \frac{\left \vert \fourier{q}{a_i} \right \vert^2}{\energyman{q} \left(\energyman{q}-E\right)} \\
\nonumber &=& - C^{-1} g^4(\mu_i) \left \vert \psi(a_i) \right \vert^2 \int_p \frac{\left \vert \fourier{p}{a_i} \right \vert^2}{\energyman{p} \left(\energyman{p}-E\right)} \\
\nonumber &=& - C^{-1} g^4(\mu_i) \left \vert \psi(a_i) \right \vert^2 \int_p \left \vert \fourier{p}{a_i} \right \vert^2 \left( \frac{1}{\left(\energyman{p}-E\right)^2} -\frac{E}{\energyman{p} \left(\energyman{p}-E\right)^2}\right) \\
&=& - <\Psi \vert H_0 \Psi>+E
\end{eqnarray}
In going from the second line of this equation to the third, the relation in eq.~(\ref{consistencyeq}) appropriate for the case of a single Dirac-delta center was used. Similarly, one obtains the last line from the previous one by substituting the value of the normalization constant in the single Dirac-delta center case. The interaction energy is also infinite and delicately cancels the infinity in the relativistic energy. A state with infinite relativistic energy is not actually in the domain of the non-interacting part of the Hamiltonian. However, the self-adjoint extension of the Hamiltonian extends the domain of the Hamiltonian to include such eigenstates, which would otherwise be excluded, with finite (total) energy. Conclusions regarding the relativistic and interaction energies carry over to the many center case with no modification at all but the algebra is more complicated.

\subsection{Bound State Energies in the Tunnelling Regime}
\label{tunnellingregime}
\par In the aforementioned limit of $md_{ij}^{min}\gg 1$, which hereafter will be called the tunnelling regime, it is possible to evaluate the off-diagonal elements of the principal operator matrix for single particle states analytically up to corrections of ${\mathcal O}((md_{ij})^{-1})$. Of course, on Riemannian manifolds $d_{ij} \equiv d(a_i,a_j) $ should be interpreted as the geodesic distance between points $a_i$ and $a_j$ and $d_{ij}^{min}$ is the smallest of these geodesic distances between different points.
\begin{equation}
\label{tunnelling}
\Phi_{ij}=-\frac{1}{\sqrt{\pi}} \int_0^{\infty} e^{-s^2/4} ds \int_0^{\infty} e^{-um^2} K_u(a_i,a_j) e^{\sqrt{u}Es} du
\end{equation}
The equation above giving the off-diagonal elements on Riemannian manifolds provides our starting point. We use the following identity
\begin{equation}
\label{scaleheatkernel}
K_u(x,y;g)=\frac{1}{d_{ij}^2}K_{u/d_{ij}^2}(x,y;g\dij^{-2})
\end{equation}
This identity expresses the fact that the heat kernel is invariant under a simultaneous scaling of the distances and the time variable. However, the lhs of the equation has mass dimension $n$ in $n$-dimensions whereas the heat kernel on the rhs is dimensionless. That's the reason for the prefactor on the rhs in the above equation. It is possible to obtain this prefactor by using the stochastic completeness condition of the heat kernel given in eq.~(\ref{stochasticcomp}). We also need the small time approximation for the heat kernel on compact manifolds \cite{chavel} given by
\begin{equation}
\label{earlytimeheatkernel}
\lim_{t \rightarrow 0^+} K_t(x,y) \simeq \frac{e^{-d^2(x,y)/(4t)}}{(4\pi t)^{n/2}} \rho(d(x,y)) \left[\sum_{j=0}^k t^j c_j(x,y;g) + {\mathcal O}(t^{k+1})\right]
\end{equation}
Moreover, we need to keep only the first term in the sum since the other terms are higher order in an expansion in powers of $a^{-1}$. We arrive at the following results:
\begin{eqnarray}
\label{tunnellingheatkernel}
\nonumber \Phi_{ij}(E=\mu_i) &=& -\frac{\rho(1) c_0(a_i,a_j;g\dij^{-2})}{\sqrt{2\pi}} \cdot \frac{e^{-m\dij \sqrt{1-(\mu_i/m)^2}}}{\left(m\dij \sqrt{1-(\mu_i/m)^2}\right)^{1/2}} \times \\
&& \left \{ \begin{array}{lr} \frac{1}{2} & \left\vert \mu_i/m \right\vert \ll (m\dij)^{-1/2} \\ 1 & (m\dij)^{-1/2} \ll \mu_i/m \ll 1-(m\dij)^{-1} \end{array} \right.
\end{eqnarray}
There is an easy way to understand the parametric dependence of the exponential in the final result eq.~(\ref{tunnellingheatkernel}).
\begin{equation}
\Phi_{ij} \sim e^{ikx} \sim e^{id_{ij}\sqrt{E^2-m^2}} \sim e^{id_{ij}\sqrt{\mu_i^2-m^2}} \sim e^{-d_{ij}\sqrt{m^2-\mu_i^2}} \sim e^{-m\dij \sqrt{1-(\mu_i/m)^2}}
\end{equation}
\par If $(m\dij)^{-1/2} \ll -\mu_i/m \leq 1$ the result is, except for the exponential, parametrically different\footnote{As $b_i$ approaches zero from above and below there are transition regions where the coefficient interpolates between $1$ and $1/2$ in one case and where the parametric behavior changes smoothly in the other, so that there is no issue of non-analiticity.}. 
\begin{eqnarray}
\label{negenergytunnelresult}
\Phi_{ij}(E=\mu_i)=-\frac{\rho(1) c_0(a_i,a_j;g\dij^{-2})}{2\pi} \cdot \frac{e^{-m\dij}}{m\dij \vert \mu_i/m \vert} & (m\dij)^{-1/2} \ll -\mu_i/m < 1 &
\end{eqnarray}
The equations given above are also valid on flat surfaces with the substitution $\rho(1) c_0(a_i,a_j;g\dij^{-2})=1$. Details of this calculation can be found in the appendix. We can now use these expressions in eq.~(\ref{perturbative}) to find the change in the binding energy in the tunnelling regime due to the presence of the other Dirac-delta centers.
\par In the tunnelling regime, we know what the ground and excited states are, these are precisely the eigenvectors obtained by perturbing the eigenvectors of the diagonal part of the principal operator matrix. We also know that these states will have energy no greater than $m$ provided that $\mu_i < m$ and that the difference between the two is sufficiently large. This is because the change in the binding energy due to the other Dirac-delta centers is infinitesimally small in this regime and can not alter this inequality. However, in other regions of the parameter space, we have so far taken the existence of the ground and excited states with energy less than $m$ for granted. We will now partially bridge this gap by proving that there always exists a ground state with energy no greater than $m$.
\par First, we begin by showing an important result for the variation of the eigenvalues with energy. Eigenvalues are given by the following expectation value
\begin{equation}
\label{eigenvalueeq}
\lambda^m(E)=<\chi^m(E)\vert \Phi(E) \chi^m(E)>
\end{equation}
The states $\chi^m(E)$ are eigenvectors of the principal operator with eigenvalue $\lambda^m(E)$, not necessarily zero. At the bound state energies, (only) the eigenvalue(s) corresponding to the bound state eigenvector(s) becomes zero. According to the Feynman-Helman theorem, the change in these eigenvalues due to a change in energy is given by the following equation
\begin{equation}
\label{feynmanhelman}
\frac{\partial \lambda^m(E)}{\partial E}=<\chi^m(E)\vert \frac{\partial \Phi(E)}{\partial E} \chi^m(E)>
\end{equation}
There is no summation over $m$ in the above equations. Inserting the principal operator in eq.~(\ref{regularphi}) into the above equation we arrive at the following:
\begin{eqnarray}
\label{evaluevariation}
\nonumber \frac{\partial \lambda^m}{\partial E}&=&\frac{-1}{\sqrt{\pi}} \int_0^{\infty} s e^{-s^2/4} ds \int_0^{\infty} du \sqrt{u} e^{-um^2+sE\sqrt{u}} \sum_{i,j} \chi^{m*}_i K_u(a_i,a_j) \chi^m_j \\
\nonumber &=&\frac{-1}{\sqrt{\pi}} \int_0^{\infty}  se^{-s^2/4} ds \int_0^{\infty}  \sqrt{u} e^{-um^2+sE\sqrt{u}} du \int \! d_gz \left \vert \sum_i K_{u/2}(a_i,z) \chi_i^m \right \vert^2 \\
& < & 0
\end{eqnarray}
In going from the first line to the second we have used the reproducing and symmetry properties of the heat kernel as well as the fact that the heat kernel is real. As we will see, what is important for our problem is not the sign of the derivative but that it has the same sign for all eigenvalues.
\par According to the Cauchy interlacing theorem \cite{bhatia}, if the last row and column of an $(N+1) \times (N+1)$ matrix is deleted, the eigenvalues of the resulting matrix $\lambda^\prime(E)$ are interlaced by the eigenvalues of the original one $\lambda(E)$ in the following manner:
\begin{equation}
\label{cauchyinterlace}
\lambda_1(E) \leq \lambda^\prime_1(E) \leq \lambda_2(E) \leq \lambda^\prime_2(E) \leq \cdots \leq \lambda^\prime_N(E) \leq \lambda_{N+1}(E)
\end{equation}
If we begin with a single Dirac-delta center with binding energy $\mu_1^\prime \leq m$ then adding a second Dirac-delta center, according to the Cauchy interlacing theorem, produces two eigenvalues which satisfy the following inequality
\begin{equation}
\lambda_1(\mu_1^\prime) \leq \lambda^\prime_1(\mu_1^\prime)=0 \leq \lambda_2(\mu_1^\prime)
\end{equation}
If we then adjust the energy to $\mu_1$ to make the eigenvalue of the new matrix zero i.e. $\lambda_1(\mu_1)=0$, as a consequence of eq.~(\ref{evaluevariation}) this new binding energy satisfies $\mu_1 \leq \mu_1^\prime \leq m$ and is the new ground state energy. One can repeat the same argument to conclude that there will be a ground state when another Dirac-delta center is added and so on. This completes the proof that there always is a ground state with energy not exceeding the mass of the particle. The same argument would have applied had the sign in eq.~(\ref{evaluevariation}) been reversed but the ground state energy would have been $\mu_2$ satisfying $\lambda_2(\mu_2)=0$. The third case in which the eigenvalues do not change with energy corresponds to a principal operator that is identically zero and is, therefore, irrelevant.
\par Perron-Frobenius theorem \cite{roger} from matrix analysis guarantees that the ground state with energy less than $m$ whose existence was proved by the Cauchy interlacing theorem is the unique ground state. According to this theorem, a matrix $A \in {\mathbb R}^{n\times n}$ whose elements are all positive has an eigenvalue $\lambda_{PF}$ that is real and positive, with positive left and right eigenvectors, for any other eigenvalue $\lambda$ we have $\lambda_{PF} > \vert \lambda \vert$ and the eigenvalue $\lambda_{PF}$ is simple that is it has multiplicity one.
\par The reality property is trivially satisfied in our case, since the principal operator matrix is hermitian, however it does not fit the description in the theorem above. In spite of this, it can be rendered positive as follows:
\begin{equation}
\label{matrixsubtract}
M\equiv -\left(\Phi-\Phi_{ii}^{max}(E_*){\mathbb I}\right)
\end{equation}
Here $E_*$ is the lower bound for the ground state energy. According to the Perron-Frobenius theorem, the matrix $M$ has a unique eigenvalue that is greater than all other eigenvalues, this implies that the matrix $\left(\Phi-\Phi_{ii}^{max}(E_*){\mathbb I}\right)$ has a corresponding eigenvalue which is more negative than all the other eigenvalues and is unique. This eigenvalue flows to $-\Phi_{ii}^{max}(E_*)$ as the energy is changed. During the flow it retains its uniqueness and becomes a bound state when $\lambda_{PF}=-\Phi_{ii}^{max}(E_*)$ since the eigenvalues were all shifted down by this value. This state is the ground state because all the other eigenvalues intersect $-\Phi_{ii}^{max}(E_*)$ at a higher value of energy due to the negative sign in eq.~(\ref{evaluevariation}). This completes the proof that the ground state is unique. 
\par Moreover, the fact that the elements of the left and right eigenvectors, which in our case are the same, can all be chosen positive means that the ground state has no nodes just as in elementary quantum mechanics. This is because the elements of the eigenvectors correspond to $R(a_i)$ in eq.~(\ref{wavefuncheatkernel}) and when these are all positive the wavefunction is positive over the whole manifold.

\section{Lower Bounds on the Ground State Energy}
\label{groundbound}
The energy of the ground state should neither be too high nor too low so as not to violate our approximation of neglecting pair formation due to the interaction with the source. We have proved, in the previous section, that there always exists a ground state with energy not exceeding the boson mass. In this section, we will show that it is also possible to constrain the ground state energy from below by adjusting the parameters in the problem like the interspacing between the delta-function centers, their number and the minimum binding energy to a single Dirac-delta center.
\par In order to place lower bounds on the ground state we will use a theorem from matrix analysis. Ger\u{s}gorin theorem \cite{roger} states that the eigenvalues of the renormalized principal operator matrix $\Phi$ are located within the union of $N$ discs
\begin{equation}
\label{gersgorin}
\bigcup_{i=1}^N \left \vert \Phi_{ii}(E)-\lambda_i \right \vert \leq \sum_{j=1 \atop j \neq i}^N \left \vert \Phi_{ij}(E) \right \vert
\end{equation}
where the eigenvalues are $\lambda_i=0$ in our problem. If we do not want to have any solutions below a given value of energy $E_*$ then the above inequality should not be satisfied for any $i$. This leads to the following inequality between the minimum on-diagonal and maximum off-diagonal matrix elements of the principal operator matrix
\begin{equation}
\label{violationgersgorin}
\left \vert \Phi_{ii}(E_*) \right \vert \geq \left \vert \Phi_{ii}(E_*) \right \vert^{min} > \left( N-1\right) \left \vert \Phi_{ij}(E_*) \right \vert^{max} \geq \sum_{j=1 \atop j \neq i}^N \left \vert \Phi_{ij}(E_*) \right \vert
\end{equation}
The key point here is that once we find such an $E_*$, then the absence of solutions is guaranteed for $E \leq E_*$. This is because the absolute value of the off-diagonal elements of the heat kernel decrease with decreasing energy for all energies whereas the absolute value of the on-diagonal ones increase for $E \leq \mu_i^{min}$. The ground state $E_{gr}$ will then be restricted to the range $E_{gr} \geq E_*$.

\subsection{Flat Space ${\mathbb R}^2$}
\label{groundboundflat}
We found the matrix elements of the principal operator on flat spaces. They were given by the following formulas:
\begin{eqnarray}
\label{phiflatoffdiag}
\nonumber \Phi_{ii} &=& \frac{1}{2\pi} \ln \left(\frac{m-E}{m-\mu_i}\right) \\
\Phi_{ij} &=& \frac{-1}{2\pi}\int_0^{\infty}\frac{ds}{\sqrt{s^2+1}} \mbox{ exp} \left[ -d_{ij} \left( m\sqrt{s^2+1}-Es\right) \right]
\end{eqnarray}
In order to place an upper bound on this integral we use the following relation
\begin{eqnarray}
\label{flatphioffdiaglow}
\nonumber \vert \Phi_{ij} \vert & \leq & \frac{1}{2\pi}\int_0^{\infty} e^{-d_{ij}(m-E)s} ds \\
& \leq & \frac{1}{2\pi d_{ij}(m-E)}
\end{eqnarray}
Therefore, there will be no solution to the eigenvalue equation for values of energy below $\Emin$ for which the following bound is satisfied
\begin{eqnarray}
\label{flatphiineq}
\frac{m-E_*}{m-\mumin}>W\left(\frac{(N-1)}{\dijmin \left(m-\mumin \right)}\right)
\end{eqnarray}
where $\mu_i^{min}$ is the smallest of the binding energies to single Dirac-delta function centers and $W(z\ln (z)) \equiv z$ is the Lambert W-function, also called the omega or the product-log function\footnote{The limiting forms of $W(z)$ are: $W(z) \simeq z+1$ for $z \ll 1$ and $W(z) \simeq \frac{z}{\ln z}\left( 1+\frac{\ln \ln z}{\ln z}\right)$ for $z \gg 1$. It is defined only for $z \geq 0$.}. The corresponding lower bound for the ground state energy $E_{gr}$ is
\begin{eqnarray}
E_{gr} \geq m-\left(m-\mumin \right)W\left(\frac{(N-1)}{\dijmin \left(m-\mumin \right)}\right)
\end{eqnarray}
This result shows that for a fixed number $N$ of Dirac-delta centers it is possible by choosing the $\dijmin$ properly, to raise the ground state energy above a critical value so that the neglect of pair creation is indeed justified.

\subsection{Compact Riemannian Manifolds}
\label{groundboundcompact}
\par The compact Riemannian manifolds over which we will investigate the issue of a lower bound on the ground state energy are those that have a Ricci curvature greater than or equal to zero everywhere. The $n$-dimensional sphere ${\mathbb S}^n$ provides a well known example to such manifolds. We need to constrain the on-diagonal elements of the principal operator matrix from below and those of the off-diagonal ones from above. In order to achieve this, we look for upper and lower bounds on the off- and on-diagonal elements of the heat kernel, respectively. Below is the first corollary \cite{wang} that we will need.
\par \textbf{Corollary 1:} Let $\mathcal M$ be a compact manifold. Suppose that the Ricci curvature of $\mathcal M$ satisfies $Ric_{\mathcal M} \geq -K, K\geq 0$. Then $\forall u>0$ and $x \in \mathcal M$
\begin{eqnarray}
\label{compactdiagupp}
K_u(x,x) \leq \frac{1}{V(\mathcal M)}+A'u^{-n/2}
\end{eqnarray}
where $n= \mbox{dim } {\mathcal M}$, $A' \equiv A'(d,K,V({\mathcal M}))>0$ and $V({\mathcal M})$ is the volume of the manifold and $d$ its diameter.
\par The following corollary \cite{grigoryan} constrains the off-diagonal elements of the heat kernel from above.
\par \textbf{Corollary 2}: Assume that for some points $x,y \in {\mathcal M} \mbox{   and   } \forall u>0,$
\begin{eqnarray}
\label{func}
K_u(x,x) \leq \frac{C'}{f(u)} \mbox{   and   } K_u(y,y) \leq \frac{C'}{g(u)}
\end{eqnarray}
where $f$ and $g$ are increasing positive functions on $(0,+\infty )$ satisfying the regularity condition given below. Then, $\forall u>0, D>2$ and for some $\epsilon >0$
\begin{eqnarray}
\label{offdiagupp}
K_u(x,y)\leq \frac{C}{\sqrt{f(\epsilon u) g(\epsilon u)}} \ \mbox{exp} \left(-\frac{d^2(x,y)}{2Du}\right)
\end{eqnarray}
\textit{regularity condition:} There are numbers $B>1$ and $b>1$ such that
\begin{eqnarray}
\label{regularity}
\frac{f(bs)}{f(s)} \leq B \frac{f(bu)}{f(u)}
\end{eqnarray}
$\forall \mbox{  } 0<s<u$ and $C=C(b,B)$, $\epsilon=\epsilon(b,B)$.
\par By inspection of eqs.~(\ref{compactdiagupp}) and (\ref{func}), one sees that these functions can be chosen as
\begin{eqnarray}
\label{regularfunc}
f(u)=g(u)=\left(\frac{1}{V(\mathcal M)}+A'u^{-n/2}\right)^{-1}
\end{eqnarray}
with $C'=1$. It is obvious that these functions are positive and one can easily verify that they are increasing. They satisfy the regularity condition (\ref{regularity}) with $B=b^{n/2}$. Therefore, we can apply Corollary 2 to conclude that the upper bound for the off-diagonal elements of the heat kernel is
\begin{eqnarray}
\label{compactoffdiagupp}
K_u(x,y)\leq C(\epsilon)\left(\frac{1}{V(\mathcal M)}+A(\epsilon)(4\pi u)^{-n/2}\right) \mbox{exp} \left(-\frac{d^2(x,y)}{2Du}\right)
\end{eqnarray}
In order to place a lower bound on the diagonal elements of the heat kernel we make use of the following theorem \cite{davies}.
\par \textbf{Theorem 1:} ${\mathcal M}$ is a complete Riemannian manifold with $Ric \geq 0$. We have 
\begin{eqnarray}
\label{compactlow}
K_u(x,y) \geq (4\pi u)^{-n/2} \exp\left[-d^2(x,y)/(4u)\right]
\end{eqnarray}
$\forall x,y \in {\mathcal M}$ and $u>0$.
Therefore, we find the lower bound to be
\begin{eqnarray}
\label{compactdiaglow}
K_u(x,x) \geq (4\pi u)^{-n/2} 
\end{eqnarray}
\par Plugging these expressions in the integrals for the diagonal and off-diagonal matrix elements of the principal operator gives the following bounds
\begin{eqnarray}
\label{phicompactbounds}
\nonumber \Phi_{ii} & \geq & \frac{1}{2\pi} \ln \left(\frac{m-E}{m-\mu_i}\right) \\
\nonumber \vert \Phi_{ij} \vert & \leq & \frac{C(\epsilon)}{2\pi} \left[ A(\epsilon)-\frac{2\pi}{mV({\mathcal M})} \cdot \frac{\partial}{\partial m}\right] \left(\frac{\sqrt{D/2}}{\dij (m-E)}\right) \\
& \leq & \frac{C(\epsilon)}{2\pi} \left[ A(\epsilon)+\frac{2\pi}{m(m-E)V({\mathcal M})} \right] \left(\frac{\sqrt{D/2}}{\dij (m-E)}\right)
\end{eqnarray}
Combining these gives us the following inequality that needs to be satisfied for a solution to the eigenvalue equation not to exist
\begin{eqnarray}
\label{compactphiineq}
\left(\frac{m-E_*}{m-\mumin}\right)^2>W\left({\frac{4C(\epsilon)(N-1)\sqrt{D/2}}{m\dijmin (m-\mumin)^2 V({\mathcal M})} \left[A(\epsilon) m^2 V({\mathcal M})+\pi \right]}\right)
\end{eqnarray}
In obtaining the inequality given above, we have replaced $m-E$ multiplying $A(\epsilon)$ by $2m$ since $E > -m$ and this only gives a more stringent constraint than the previous one. So the ground state energy satisfies the following inequality
\begin{eqnarray}
E_{gr} \geq m-\left(m-\mumin \right)W^{1/2}\left({\frac{4C(\epsilon)(N-1)\sqrt{D/2}}{m\dijmin (m-\mumin)^2 V({\mathcal M})} \left[A(\epsilon) m^2 V({\mathcal M})+\pi \right]}\right)
\end{eqnarray}
\par On compact manifolds, there may be more than one distance minimizing curve connecting distinct points. Away from a fixed point, the set of points for which this is the case are called the cut loci of that point. The on- and off-diagonal bounds for the heat kernel given in this section are valid even if the Dirac-delta centers fall on the cut loci of one another.

\subsection{Cartan-Hadamard Manifolds}
\label{groundboundcartanhadamard}
\par A manifold ${\mathcal M}$ is called a Cartan-Hadamard manifold \cite{grigoryan3} if ${\mathcal M}$ is a geodesically complete, simply connected, non-compact Riemannian manifold with non-positive sectional curvature everywhere. The $n$-dimensional flat ${\mathbb R}^n$ and hyperbolic spaces ${\mathbb H}^n$ are the best known examples of Cartan-Hadamard manifolds. The mass of the particle on such manifolds is shifted from its value on compact and flat surfaces, due to the geometry of the manifold, to $\sqrt{m^2+\lambda_{min}}$. On Cartan-Hadamard manifolds, the on-diagonal elements of the heat kernel are constrained from below in the manner given below \cite{grigoryan3}.
\par \textbf{Theorem 3:} Assume that the sectional curvature inside the ball $B(o,r)$ is bounded below by $-K^2_{max}(r)$. Then, for all $u>0$, $x\in {\mathcal M}$ and $\delta>0$,
\begin{equation}
\label{hadamarddiaglow}
K_u(x,x) \geq \frac{c}{(4\pi u)^{n/2}} \mbox{ exp} \left[-(\lambda_1({\mathcal M})+\delta)u \right]
\end{equation}
where $c \equiv c(o,\delta) >0$. The minimum eigenvalue of the Laplacian $\lambda_1({\mathcal M})$ is restricted to the following range \cite{carmo, chavel}
\begin{equation}
\frac{1}{4} (n-1)^2 K_{max}^2 \geq \lambda_1({\mathcal M}) \geq \frac{1}{4} (n-1)^2 K_{min}^2
\end{equation}
for a Cartan-Hadamard manifold whose sectional curvature is bounded from above by $-K_{min}^2$.
\par Here is an important theorem \cite{hoffman} regarding Cartan-Hadamard manifolds. 
\par \textbf{Theorem 4:} Any Cartan-Hadamard manifold ${\mathcal M}$ of the dimension $n$ admits the isoperimetric function $I(v)=\kappa v^{\frac{n-1}{n}}$, $\kappa>0$.
\par Another theorem \cite{grigoryan2} that we need is below.
\par \textbf{Theorem 5:} Assume that manifold ${\mathcal M}$ admits a non-negative continuous isoperimetric function $I(v)$ such that $I(v)/v$ is non-increasing. Let us define the function $f(u)$ by
\begin{equation}
\label{function}
u=4\int_0^{f(u)} \frac{vdv}{I^2(v)}
\end{equation}
assuming the convergence of the integral in the above equation at 0. If the function f satisfies in addition the regularity condition (\ref{regularity}) then, $\forall x,y\in {\mathcal M},u>0,D>2$ and some $\epsilon >0$,
\begin{equation}
K_u(x,y) \leq \frac{C'}{f(\epsilon u)} \mbox{ exp} \left(\frac{-d^2(x,y)}{2Du}\right )
\end{equation}
\par The isoperimetric function is non-negative and the function obtained from eq.(\ref{function})
\begin{equation}
f(u)=\left(\frac{\kappa^2}{2n}u\right)^{n/2}
\end{equation}
meets all the requirements of this theorem including the regularity condition (\ref{regularity}) (with any $B>1$ in 2 dimensions). Hence, the upper bound on the off-diagonal elements of the heat kernel on Cartan-Hadamard manifolds is given by
\begin{equation}
\label{hadamardoffdiagupp}
K_u(x,y) \leq \frac{C(\epsilon ,\kappa)}{(4\pi u)^{n/2}} \mbox{ exp} \left(\frac{-d^2(x,y)}{2Du}\right )
\end{equation}
The bounds on the on and off- diagonal elements of the heat kernel lead to the following inequalities for the matrix elements of the principal operator
\begin{eqnarray}
\label{phihadamardbounds}
\nonumber \Phi_{ii} & \geq & \frac{c(\delta)}{2\pi} \ln \left(\frac{m_{CH}-E}{m_{CH}-\mu_i}\right) \\
\vert \Phi_{ij} \vert & \leq & \frac{C(\epsilon ,\kappa)\sqrt{D/2}}{2\pi \dij (m-E)}
\end{eqnarray}
where
\begin{equation}
\label{mcartan}
m_{CH}^2=m^2+\frac{1}{4}K_{max}^2+\delta(K^2)
\end{equation}
Combining these gives us the following inequality that needs to be satisfied for a solution to the eigenvalue equation not to exist
\begin{eqnarray}
\label{hadamardphiineq}
\frac{m-E_*}{m_{CH}-\mumin}>W\left(\frac{(N-1)C(\epsilon ,\kappa) \sqrt{D/2}}{c(\delta) \dijmin \left(m_{CH}-\mumin \right)}\right)
\end{eqnarray}
We replaced $m_{CH}$ in the numerator on the lhs by $m$ as $m_{CH} \geq m$ and this replacement only results in a more stringent constraint than the one before. Then the inequality for the ground state becomes
\begin{eqnarray}
E_{gr} \geq m-\left(m_{CH}-\mumin\right)W\left(\frac{(N-1) C(\epsilon ,\kappa) \sqrt{D/2}}{c(\delta) \dijmin \left(m_{CH}-\mumin \right)}\right)
\end{eqnarray}
\par Using the large $z$ approximation of the Lambert W-function, one can check in the perturbative limit, that the ground state energy on flat and compact manifolds approaches $m$ logarithmically from below. However, on Cartan-Hadamard manifolds, the bound we obtain for the ground state in the perturbative limit is not as stringent. We could only prove that $E_{gr} \geq m$ whereas on such manifolds we would expect $E_{gr} \geq \sqrt{m^2+\lambda_1({\mathcal M})} \geq m$.

\section{Conclusion}
\label{conclusion}
In this paper, we utilized the orthofermion algebra technique to convert the problem of determining the spectrum of relativistic bosons interacting with $N$ localized sources to one where we had to solve for the eigenstates of the finite dimensional principal operator matrix with vanishing eigenvalue. The Delta-function type Hamiltonian that we are interested in was not well-defined and hence would lead to a principal operator matrix that was not well-defined either. We rendered the original Hamiltonian well-defined by the use of the heat kernel which approaches a Dirac-delta function in the limit $\epsilon \rightarrow 0^+$. \par The principal operator was found using this Hamiltonian and it was noticed that in order to get physical results that did not depend on the cut-off, the coupling constant had to depend on the cut-off in a particular manner. After regularizing the principal operator, we chose the renormalization condition and made the natural choice for which the bound state energy in the case of a single Dirac-delta function center case coincided with the experimentally measured binding energy to that center. 
\par We could not find the eigenstates of the principal operator and the bound state energies corresponding to them in the general case. However, in the tunnelling regime, we managed to find the eigenstates and the bound state energies on 2 dimensional compact manifolds and flat surfaces. Also, we found the wavefunctions for single particle states on general manifolds and determined that multiparticle states could be written as products of these single particle states.
\par We made use of the Cauchy interlacing theorem to prove that a ground state with energy not exceeding the boson mass exists so long as the single Dirac-delta center binding energies are not above this mass. Most importantly, we proved that the ground state energy has a lower bound on flat, compact and Cartan-Hadamard type manifolds for appropriate choices of the number of Dirac-delta centers $N$, their separations $\dij$ and the initial binding energies $\mu_i$ to single Dirac-delta centers. Thus, validating the neglect of particle production due to the interaction with the source, on a large class of 2-dimensional Riemannian manifolds.
\par We would like to extend the proof that the ground state with energy no greater than the boson mass exists to excited states or to show under what conditions excited states that meet this requirement exist in future research. It would also be interesting to find approximate/exact solutions in regimes where the bound state energy is fairly different from the binding energy to the single Dirac-delta center.

\section{Acknowledgements}
\label{acknowledge}
O.T. Turgut likes to thank F. Erman and B.T. Kaynak for helpful discusssions. This research is supported by the Scientific and Technological Research Council of Turkey (T\"{U}B\.{I}TAK) under the National Postdoctoral Research Scholarship Program 2218 (2218 - Yurt \.{I}\c{c}i Doktora Sonras\i\ Ara\c{s}t\i rma Burs Program\i).

\section{Appendix: Calculation of Bound State Energies in the Tunnelling Regime}
\label{appendix}
In this appendix, we give those details that were omitted in the main body of the text related to the calculation of bound state energies in the tunnelling regime. We first consider the case $a^{-1/2} \ll b_i \ll 1-a^{-1}$ where $a \equiv m\dij$ and $b_i \equiv \mu_i/m$. We start with eq.~(\ref{tunnelling}),
\begin{eqnarray}
\label{offdiag}
\nonumber \Phi_{ij} &=& -\frac{1}{\sqrt{\pi}} \int_0^{\infty} e^{-s^2/4} ds \int_0^{\infty} e^{-um^2} K_u(a_i,a_j) \ e^{\sqrt{u}Es} \ du \\
\nonumber &=& -\frac{1}{\sqrt{\pi}} \int_0^{\infty} ds \int_0^{\infty} du \ e^{-um^2} K_u(a_i,a_j) \ e^{-\left(s-2\sqrt{u}E\right)^2/4} \ e^{uE^2} \\
\nonumber &=& -\frac{1}{\sqrt{\pi}} \int_0^{\infty} ds \int_0^{\infty} du \ K_u(a_i,a_j) \ e^{-\left(s-2\sqrt{u}E\right)^2/4} \ e^{-um^2\left(1-b^2\right)}, b\equiv \frac{E}{m}, \\
\nonumber &=& -\frac{2}{\sqrt{\pi}} \int_0^{\infty} du \ K_u(a_i,a_j) \ e^{-um^2\left(1-b^2\right)} \int_{-\sqrt{u}E}^{\infty} e^{-s^2} ds, \ u\equiv \frac{u^{\prime}}{A}, \\
\nonumber &=& -\frac{2}{\sqrt{\pi}} \int_0^{\infty} \frac{du^{\prime}}{A} K_{u^{\prime}/A}(a_i,a_j) \ e^{-u^{\prime}m^2\left(1-b^2\right)/A} \int_{-\frac{\sqrt{u^{\prime}}mb}{A}}^{\infty} e^{-s^2} ds
\end{eqnarray}
then we plug the following equation in the one above
\begin{eqnarray}
\label{scaling}
\nonumber K_{u}(x,y;g) = \frac{1}{\dij^2} K_{u/\dij^2}(x,y;g\dij^{-2})
\end{eqnarray}
and we get
\begin{eqnarray}
\nonumber \Phi_{ij} &=& -\frac{2}{\sqrt{\pi}} \int_0^{\infty} \frac{du}{A} \cdot \frac{1}{\dij^2} K_{u/\left(A\dij^2\right)}\left(a_i,a_j;g\dij^{-2}\right) \ e^{-u\left(1-b^2\right)m^2/A} \int_{-\frac{\sqrt{u}mb}{A}}^{\infty} e^{-s^2} ds \\
\nonumber &=& -\frac{2}{\sqrt{\pi}} \int_0^{\infty} du \ (m\dij)^{-1} \ K_{u/\left(m\dij\right)}\left(a_i,a_j;g\dij^{-2}\right) \ e^{-um\dij\left(1-b^2\right)} \int_{-b\sqrt{um\dij}}^{\infty} e^{-s^2} ds
\end{eqnarray}
where we have chosen $A \equiv \frac{m}{\dij}$. At this point, we argue that the $s$ integral is independent of $u$ provided $b\sqrt{ua} \gg 1$. This will be the case if the $u$ integral is dominated by $u$'s of ${\mathcal O}(1)$. This turns out to be case as can be checked a posteriori. The other is a necessary condition that $\mu_i$ must satisfy in order for this method to work. Another condition that has to be satisfied in order for this to be a good estimate to the integral is that the exponential be highly damped that is $a \left(1-b^2\right) \sim a(1-b) \gg 1$ for $b \simeq 1$. For values of $u$ that dominate the integral, the time argument of the heat kernel is a small number and the small time approximation for the heat kernel given below may be used.
\begin{eqnarray}
\label{earlytimeheatkernel}
\nonumber \lim_{t \rightarrow 0^+} K_t(x,y) \simeq \frac{e^{-d^2(x,y)/(4t)}}{(4\pi t)^{n/2}} \rho(d(x,y)) \left[\sum_{j=0}^k t^j c_j(x,y;g) + {\mathcal O}(t^{k+1})\right]
\end{eqnarray}
Once the small time approximation is written down it can be seen that only the first term should be kept since the higher order terms are suppressed by powers of $a^{-1}$. We continue with the following equation after having taken the above simplifications into account
\begin{eqnarray}
\nonumber \Phi_{ij} &\simeq& -\frac{2\rho}{\sqrt{\pi}} \int_0^{\infty} du \ a^{-1} \ \frac{e^{-1/\left(4u/a\right)}}{4\pi\left(u/a\right)} \ e^{-ua\left(1-b^2\right)} \int_{-\infty}^{\infty} e^{-s^2} ds \\
\nonumber &=& -\frac{\rho}{2\pi} \int_0^{\infty} \exp\left\{ -a\left[u\left(1-b^2\right)+\frac{1}{4u}\right]\right\} \ \frac{du}{u}
\end{eqnarray}
The huge damping of the exponential allows us to make a saddle point approximation. We need the value of the function and its second derivative at the point where it is stationary
\begin{eqnarray}
\nonumber f(u) &=& u\left(1-b^2\right) + \frac{1}{4u}, \\
\nonumber f^{\prime}(u) &=& 1-b^2-\frac{1}{4u^2}, \\
\nonumber f^{\prime \prime}(u) &=& +\frac{1}{2u^3}>0, \\
\nonumber f^{\prime}\left(u_0\right) &=& 1-b^2-\frac{1}{4u_0^2} = 0 \rightarrow u_0= \frac{1}{2\sqrt{1-b^2}},
\end{eqnarray}
using these in the equation above we find the following
\begin{eqnarray}
\nonumber \Phi_{ij} &\simeq& -\frac{\rho(1) c_0(a_i,a_j;g\dij^{-2})}{2\pi} \int_0^{\infty} \exp\left\{-a\left[f\left(u_0\right)+\left(u-u_0\right)f^{\prime}\left(u_0\right)+\frac{1}{2}\left(u-u_0\right)^2f^{\prime \prime}\left(u_0\right)\right]\right\} \frac{du}{u} \\
\nonumber &=& -\frac{\rho(1) c_0(a_i,a_j;g\dij^{-2})}{2\pi u_0} \exp\left[-af\left(u_0\right)\right] \int_0^{\infty} \exp\left[-\frac{1}{2}a\left(u-u_0\right)^2f^{\prime \prime}\left(u_0\right)\right] du \\
\nonumber &=& -\frac{\rho(1) c_0(a_i,a_j;g\dij^{-2})}{2\pi u_0} \cdot \sqrt{\frac{2}{af^{\prime \prime}\left(u_0\right)}} \cdot \exp\left[-af\left(u_0\right)\right] \int_{-u_0\sqrt{\frac{af^{\prime \prime}\left(u_0\right)}{2}}}^{\infty} e^{-u^2} du \\
\nonumber &=& -\frac{\rho(1) c_0(a_i,a_j;g\dij^{-2})}{\pi} \cdot \sqrt{\frac{u_0}{a}} \cdot \exp\left(-a\sqrt{1-b^2}\right) \int_{-\sqrt{a}\cdot\frac{1}{2\sqrt{u_0}}}^{\infty} e^{-u^2} du \\
\nonumber &\simeq& -\frac{\rho(1) c_0(a_i,a_j;g\dij^{-2})}{\pi \sqrt{2}} \cdot \frac{\exp\left(-a\sqrt{1-b^2}\right)}{\left(a\sqrt{1-b^2}\right)^{1/2}} \int_{-\infty}^{\infty} e^{-u^2} du \\
\nonumber &=& -\frac{\rho(1) c_0(a_i,a_j;g\dij^{-2})}{\sqrt{2\pi}} \cdot \frac{\exp\left(-a\sqrt{1-b^2}\right)}{\left(a\sqrt{1-b^2}\right)^{1/2}}
\end{eqnarray}
which is eq.~(\ref{tunnellingheatkernel}) if we make the identification $b=b_i$. The case $\left \vert b_i \right \vert \ll a^{-1/2}$ can be calculated in the same way except that the $s$ integral gives a different factor. The case $-b \gg a^{-1/2}$ requires a different approach than the previous one. We begin again with eq.~(\ref{tunnelling})
\begin{eqnarray}
\nonumber \Phi_{ij} &=& -\frac{1}{\sqrt{\pi}} \int_0^{\infty} e^{-s^2/4} ds \int_0^{\infty} e^{-um^2} K_u(a_i,a_j) \ e^{\sqrt{u}Es} \ du, \ t\equiv \frac{\sqrt{u}}{\dij}s, \\
\nonumber &=& -\frac{1}{\sqrt{\pi}} \int_0^{\infty} e^{-m\dij \vert E/m\vert t} \ \dij \ dt \int_0^{\infty} K_u(a_i,a_j) \ e^{-um^2} \ e^{-\dij^2 t^2/(4u)} \ \frac{du}{\sqrt{u}}, \ u\equiv \frac{u^{\prime}}{A}, \\
\nonumber &=& -\frac{1}{\sqrt{\pi}} \int_0^{\infty} e^{-a \vert b\vert t} \ dt \int_0^{\infty} \frac{1}{\dij^2} K_{u^{\prime}/\left(A\dij^2\right)}(a_i,a_j;g\dij^{-2}) \ e^{-u^{\prime}m^2/A} \ e^{-\dij^2 t^2/(4u^{\prime}/A)} \ \dij \cdot \frac{du^{\prime}}{A} \cdot \frac{\sqrt{A}}{\sqrt{u^{\prime}}} \\
\nonumber &=& -\frac{1}{\sqrt{a\pi}} \int_0^{\infty} e^{-a \vert b\vert t} \ dt \int_0^{\infty} K_{u/a}(a_i,a_j;g\dij^{-2}) \ e^{-ua} \ e^{-a t^2/(4u)} \ \frac{du}{\sqrt{u}} \\
\nonumber &\simeq& -\frac{\rho(1) c_0(a_i,a_j;g\dij^{-2})}{\sqrt{a\pi}} \int_0^{\infty} e^{-a \vert b\vert t} \ dt \int_0^{\infty} \frac{e^{-1/\left(4u/a\right)}}{4\pi\left(u/a\right)} \ e^{-ua} \ e^{-a t^2/(4u)} \ \frac{du}{\sqrt{u}} \\
\nonumber &=& -\frac{\rho(1) c_0(a_i,a_j;g\dij^{-2}) \sqrt{a}}{2\pi} \int_0^{\infty} e^{-a \vert b\vert t} \ dt \left\{\frac{1}{2\sqrt{\pi}} \int_0^{\infty} \exp\left\{-a\left[u+\frac{\left(1+t^2\right)}{4u}\right]\right\} \ \frac{du}{u^{3/2}}\right\} \\
\nonumber &=& -\frac{\rho(1) c_0(a_i,a_j;g\dij^{-2}) \sqrt{a}}{2\pi} \int_0^{\infty} e^{-a \vert b\vert t} \cdot \frac{\exp\left[-\sqrt{a}\cdot\sqrt{a\left(1+t^2\right)}\right]}{\sqrt{a\left(1+t^2\right)}} dt \\
\nonumber &=& -\frac{\rho(1) c_0(a_i,a_j;g\dij^{-2})}{2\pi} \int_0^{\infty} \exp\left[-a\left(\vert b\vert t+\sqrt{1+t^2}\right)\right] \ \frac{dt}{\sqrt{1+t^2}}
\end{eqnarray}
Since this exponential is highly damped the major contribution will be from the $t \approx 0$ region where we can ignore the $t^2$ term in the exponential and in the denominator. These simplifications lead to the elementary integral and the result given below
\begin{eqnarray}
\nonumber \Phi_{ij} &=& -\frac{\rho(1) c_0(a_i,a_j;g\dij^{-2})}{2\pi} \cdot e^{-a} \int_0^{\infty} e^{-a\vert b\vert t} dt \\
\nonumber &=& -\frac{\rho(1) c_0(a_i,a_j;g\dij^{-2})}{2\pi} \cdot \frac{e^{-a}}{a\vert b\vert}
\end{eqnarray}
which is eq.~(\ref{negenergytunnelresult}) with $b=b_i$.

\end{document}